\documentclass[12pt]{article}
\usepackage{amssymb}
\usepackage[mathscr]{eucal}
\usepackage{citesort}
\usepackage{url}
\usepackage{a4}
\usepackage{cite}
\usepackage{graphicx}
\usepackage{psfrag}
\usepackage{subfigure}
\usepackage{latexsym}
\usepackage{amsmath}
\newcounter{mnotecount}[section]

\begin{document}
\newcommand{\g}{$\bf \bar g g^{\alpha\beta}$}
\title{Celestial mechanics of elastic bodies}

 \author{Robert Beig\\Institut f\"ur Theoretische Physik der Universit\"at
Wien\\ Boltzmanngasse 5, A-1090 Vienna, Austria\\[1cm] Bernd G. Schmidt\\
Max-Planck-Institut f\"ur Gravitationsphysik\\ Albert-Einstein-Institut\\
Am M\"uhlenberg 1, D-14476 Golm, Germany}
\maketitle

\begin{abstract}
We construct time independent configurations of two gravitating
elastic bodies. These configurations either correspond to the two
bodies moving in a circular orbit around their center of mass or
strictly static configurations.


\end{abstract}

\section{Introduction}
\label{introduction} In  the year 1933 Leon Lichtenstein published a book entitled
"Gleichgewichtsfiguren rotierender Fl\"ussigkeiten" ("equilibrium figures of rotating fluids") in
which he collected his work on this topic, published in Mathematische Zeitschrift in the years 1905
--- 1928. Lichtenstein showed existence of solutions of the nonlinear PDEs describing rotating
fluids in various configurations under the influence of Newtonian gravity. The simplest
configurations are: one axisymmetric rigidly rotating fluid ball and two self-gravitating fluid
balls in circular motion around their center of gravity.  More complicated configurations treated
are toroidal rings or even families of nested rings. All these problems have in common that the
equations become time independent in a corotating system and that they express the balance of
gravitational, centrifugal and pressure forces. Of particular importance is that the total force
and torque on the bodies must vanish. Lichtensteins method is --- in modern language --- the
implicit function theorem in the context of bifurcation theory. In one class of problems he begins
with an exact non rotating solution and finds a nearby solution with small angular velocity. More
interesting are the cases with several bodies. Starting from two point particles on a circular
orbit around their center of gravity, one "puts two small homogeneous fluid bodies on these
orbits": this is no solution to the equation, however for a small body the error is small enough to
begin an iteration which converges to an exact solution. The purpose of this paper is to show that
basically all configurations Lichtenstein treated can also be successfully studied if the bodies
are made of elastic material. The case of one slowly rotating elastic body was studied in [1]. In
this paper we treat: Firstly, one elastic body in circular motion in the gravitational field of
some central mass which is kept fixed. Secondly, two bodies in circular motion around their common
center of gravity under their mutual Newtonian gravitational field.

We also deal with a static problem, where one small body is near a non-degenerate equilibrium point
of a given gravitational field - such as that in the center of a hollow triaxial cylinder, (iv)
like (iii) - with the mutual interaction and self-interaction of both bodies taken into account.

The paper is organized as follows. Section 2 reviews elastostatics; quotes properties of the
elasticity operator and formulates the equations for general "force functionals". In section 3 we
discuss various types of forces, which in particular include centifugal and gravitational forces.
Section 4 contains the main application, the motion of  a body in an external gravitational field
and the motion of two bodies under their gravitational attraction. Section 5 is in a way separate
from the main theme of rotating bodies. It shows, that the techniques developed  can be used to
demonstrate the existence of Newtonian static two--body solutions for particular geometries of the
bodies. This possibility had always been conjectured, but the existence of  a full PDE solutions
had not been demonstrated.
\section{Elastostatics}
\label{Elastostatics}

In this section we recall the setup of nonrelativistic elastostatics. We are using the material representation where
elastic states are decribed by deformations, i.e. maps from material space, the ``body'', to physical space. A fuller and more general
description can be found in  \cite {BSR}.\\
Let $\Omega$ be a bounded open connected subset in $\mathbb{R}^3$
with smooth boundary $\partial \Omega$. The basic objects are maps
$\Phi: \Omega \rightarrow \mathbb{R}^3$. We will use $X^A$, resp.
$x^i$ for Cartesian coordinates in $\Omega$ resp. $\mathbb{R}^3$ and
$\delta_{AB}$, resp. $\delta_{ij}$ for the metrics on the domain,
resp. target space.
%We denote $\delta^A{}_i$ for the Jacobian of the identity map $\bar{\Phi}$.
There is a trivial or reference state, denoted by $\bar{\Phi}$, which we will always take to be the
identity. We assume the $\Phi$'s are sufficiently close to $\bar{\Phi}$ (see below), so that they
have a regular inverse $f:\Phi^{-1}(\Omega) \rightarrow \Omega$. We define $\Psi^A{}_i$ by the
equation
\begin{equation}\label{Psi}
\Psi^A{}_i (X) (\partial_B \Phi^i) (X) = \delta^A{}_B
\end{equation}
and the tensor $H^{AB}$ by
\begin{equation}\label{CG}
H^{AB}(X) = \Psi^A{}_i (X) \Psi^B{}_j (X) \delta^{ij}.
\end{equation}
In other words, $H^{AB}$ is the inverse of the
pulled-back-under-$\Phi$ Euclidean metric on $\mathbb{R}^3$.\\ The
material is described by the stored-energy function $\hat{w} (X,
\partial \Phi) = \hat{w}(X, H^{BC})$, smooth in its arguments and
subject to the conditions
\begin{equation} \label{stressfree}
\left(\frac{\partial \hat{w}}{\partial H^{AB}}\right)\Bigg{|}_{H^{CD} = \;\delta^{CD}}=0
\end{equation}
and that $L_{ABCD}$, defined by
\begin{equation} \label{L}
L_{ABCD} = \left(\frac{\partial^2 \hat{w}}{\partial H^{AB}\partial H^{CD}}\right)\Bigg{|}_{H^{EF} = \; \delta^{EF}},
\end{equation}
should satisfy
\begin{equation}\label{pointwise}
L_{ABCD} \;M^{AB} M^{CD} \geq C |M|^2
\end{equation}
with $C$ a positive constant and $M^{AB}=M^{(AB)}$. Condition
Eq.(\ref{stressfree}) will mean that the reference state
$\bar{\Phi}$ has vanishing stress. Condition Eq.(\ref{pointwise}),
called uniform pointwise stability, is an ellipticity condition
satisfied by standard materials. We will sometimes need the
condition that the material be homogenous and isotropic. Homogeneity
means that $\hat{w}$ is independent of $X$. Isotropy of the material
means that $\hat{w}$ depends on $H^{AB}$ only through the principal
invariants of $H^A{}_B = H^{AC} \delta_{BC}$. In the latter case
$L_{ABCD}$ is of the form
\begin{equation}\label{Liso}
L_{ABCD} = \lambda \,\delta_{AB} \delta_{CD} + 2 \mu\, \delta_{C(A} \delta_{B)D},
\end{equation}
and uniform pointwise stability is then equivalent to $\mu > 0,\,3 \lambda + 2 \mu > 0$.\\
Remark: Perfect fluids can be described by assuming the stored-energy to be a function solely of $J$, where $J$ is
the determinant of $H^A{}_B$. However in this case
condition (\ref{pointwise}) is never satisfied.\\
  The ``first Piola stress tensor''
is defined by
\begin{equation}\label{sigmahat}
\hat{\sigma}_i{}^A (X,\partial \Phi) = \frac{\partial
\hat{w}}{\partial (\partial_A \Phi^i)} = - 2 H^{AC} \Psi^B{}_i
\;\frac{\partial \hat{w}}{\partial H^{BC}}\;.
\end{equation}
Let now $\mathscr{C}$ be a neighbourhood of $\Phi = \bar{\Phi}$ in $W^{2,p}(\Omega,\mathbb{R}^3),\;p > 3$,
small enough so that each $\Phi \in \mathscr{C}$
has a $C^1$ - inverse (see p.224 of \cite{CI}).

Next let $\mathscr{E}$ be the quasilinear second-order operator sending each map $\Phi \in \mathscr{C}$ to
\begin{equation}\label{defE}
\mathscr{E}_i [\Phi](X) = (\partial_A \sigma_i{}^A) (X)
\end{equation}
where $\sigma_i{}^A(X) = \hat{\sigma}_i^A (X,\partial \Phi(X))$.
Denote $\sigma_i$ by $(\sigma_i{}^A n_A)|_{\partial \Omega}$, where
$n_A$ is the outward normal to $\partial \Omega$. Let the ``load
space'' $\mathscr{L}$ be defined by $\mathscr{L} =
W^{0,p}(\Omega,\mathbb{R}^3) \times W^{1 -
1/p,p}(\partial\Omega,\mathbb{R}^3)$. It is then well known (see
\cite{VA}), that the operator $E$ sending $\Phi \in \mathscr{C}$ to
the pair $(b = \mathscr{E}_i, \tau = \sigma_i) \in \mathscr{L}$ is
well defined and $C^1$.
Here are some properties of the map $E$.\\
{\bf{Property 1}}: $E(\bar{\Phi}) = 0$. This is obvious, since
$\hat{\sigma}_i{}^A (X,\partial \bar{\Phi})
= 0$ by Eqs.(\ref{stressfree},\ref{sigmahat}).\\
{\bf{Property 2}}: If $(b,\tau) \in E(\mathscr{C})$ then there holds
\begin{equation}\label{equil}
\int_\Omega (\xi^i\! \circ\! \Phi)(X)\; b_i (X)\; dV(X) + \int_{\partial \Omega} (\xi^i\! \circ\! \Phi)(X)\; \tau_i (X)\; dS(X) = 0,
\end{equation}
where $\xi^i(x)$ is any Killing vector of Euclidean $\mathbb{R}^3$. Note that (\ref{equil}), when
required only for translations, is independent of $\Phi$. The conditions (\ref{equil}) - often
called ``equilibration conditions'' - have the meaning of vanishing total and force
and total torque. If these conditions were violated, the body will be set into motion and no time independent solution exists.\\
Finally, the linear map $\delta E: W^{2,p}(\Omega,\mathbb{R}^3) \rightarrow \mathscr{L}$, given by the
linearization of $E$ at $\Phi = \bar{\Phi}$,
has\\
{\bf{Property 3}}: $\delta E$ is Fredholm with (6-dimensional)
kernel given by the Euclidean Killing vectors at the reference
state, i.e. $\delta \Phi^i$ of the form $\delta \Phi^i (X) =
(\xi^i\! \circ\! \bar{\Phi})(X)$. Furthermore $\delta E$ has a range
(of codimension 6) given by the subspace $\mathscr{L}_{\bar{\Phi}}$
of all $(b,\tau)$ satisfying the
equilibration conditions (\ref{equil}) at $\Phi = \bar{\Phi}$.\\
A proof of Property 3 can be found in \cite{VA}.\\
We now come to the concept of force density in the present context.
Let $F$ be a ``load map'', i.e. a  $C^1$ - map from $\mathscr{C}$ to $\mathscr{L}$. We are interested in solving equations of the form
\begin{equation}\label{equ}
E[\Phi] + \lambda F[\Phi] = 0
\end{equation}
for small $\lambda$. Our load maps will be of the form
\begin{equation}\label{load map}
F:\Phi \in \mathscr{C} \mapsto (b_i = \mathscr{F}_i[\Phi],\tau_i = 0).
\end{equation}
We are thus seeking solutions to the elastic equations with $\sigma_i$, the normal stress at the boundary, vanishing,
i.e. ``freely floating bodies''. \\
We will call a force field $\mathscr{F}$ equilibrated at $\Phi$, if $F[\Phi]$ satisfies Eq.(\ref{equil}) with $\tau_i = 0$
i.e. if
\begin{equation}\label{equil1}
\int_\Omega (\xi^i\!\circ\! \Phi)(X)\; \mathscr{F}_i [\Phi](X)
\;dV(X) = 0
\end{equation}
for all Euclidean Killing vectors $\xi$. Given a functional
$\mathscr{F}_i$, we will try to solve (\ref{equ}) by the implicit
function theorem. Suppose $\Phi_\lambda$ is a 1-parameter family of
solutions of (\ref{equ}) with $\Phi_0=\bar\Phi$. Then (\ref{equil1})
must hold for any $\lambda$, whence for $\lambda=0$. Therefore the
reference configuration must be equilibrated for the force
functional in order for being the starting solution for an
application of the implicit function theorem. Suppose we have such a
$\bar\Phi$, then the linearisation of the elasticity operator at
$\bar\Phi $ still has a non trivial finite dimensional kernel and
range. In the spirit of bifurcation theory one could try to project
the equation onto some complement of the range. Sometimes this
approach works, provided one takes "sufficiently small" but finite
bodies of any shape. We used this approach in \cite{BS1}. In the
present section we will just consider spherical bodies. Then one can
use some discrete symmetries to make the space of configurations and
loads smaller so that the implicit function theorem can be applied.
\section{Forces}
In our applications we will deal with the centrifugal force and the gravitational force under
various circumstances.
\subsection{Spatial force fields}
Both the centrifugal and an external gravitational force give rise to maps $\mathscr{F}$ of the form
\begin{equation}\label{spatial}
\mathscr{F}_i[\Phi](X) = (K_i \circ \Phi)(X),
\end{equation}
where $K_i (x)$ is a smooth vector field on  $\mathbb{R}^3$ with a potential
\begin{equation}\label{gradient}
K_i = \partial_i K
\end{equation}
satisfying
\begin{equation}\label{harmonic}
\Delta K = \delta^{ij} \partial_i \partial_j K = \mathrm{const}.
\end{equation}
We call such fields $F_i$ harmonic. It follows that $\xi^i F_i$, where $\xi^i$ is any Euclidean
Killing vector, is a harmonic function. Let us define
\begin{equation}\label{mean}
\mu_{R,x}[\xi;\Phi] =  \int_{B_R (x)} (\xi^i\! \circ\! \Phi)(X)\; (K_i\! \circ\! \Phi)(X)\;dV(x),
\end{equation}
where $B_R(x)$ is the open ball of radius $R$ centered at the point $x$.
Suppose $x_0$ is a critical  point of $K$.
Then, by the (solid form of the) mean-value theorem for harmonic functions (see \cite{GT}), there holds
\begin{equation}\label{meanbar}
\mu_{R,x_0}[\xi; \bar{\Phi}] = \int_{B_R(x_0)} \xi^i K_i \;dV(x) = |B_R|\, (\xi^i K_i)(x_0) = 0
\end{equation}
This means that $\mathscr{F}_i = K_i \circ \Phi$ is equilibrated at $\bar{\Phi}$
provided that $\Omega = B_R (x_0)$. \\
We will also use the variational formula
\begin{equation}\label{varmean}
\delta_\Phi \mu_{R,x_0}[\xi;\bar{\Phi}] \cdot \delta \Phi = \int_{B_R (x_0)} (\mathcal{L}_\xi\!
K)_i \;\delta \Phi^i \;dV(x).
\end{equation}
Taking in (\ref{varmean}) for $\delta \Phi = \eta$, with $\eta$ another Euclidean Killing vector, we find the integrand in
(\ref{varmean}) is again harmonic. Consequently
\begin{equation} \label{varmeaneta}
\delta_\Phi \mu_{R,x_0} [\xi;\bar{\Phi}] \cdot \eta = |B_R| \;(\xi^i \eta^j \partial_i \partial_j
K)(x_0)
\end{equation}
\subsection{Force fields equilibrated at all deformations}
It seems remarkable at first sight, that there should exist such force fields. Suppose there is a symmetric tensor field
$\Theta_{ij}[\Phi](x)$ on $\mathbb{R}^3$, having the property that
\begin{equation}\label{forall}
\partial_j \Theta_i{}^j = 0 \;\;\forall x \in \mathbb{R}^3 \,\backslash\; \overline{\Phi(\Omega)}.
\end{equation}
and that $\Theta_{ij}$ is of $O(|x|^{-4})$ at infinity. Then
$\mathscr{F}_i = J (\partial_j \Theta_i{}^j) \circ \Phi$ is equilibrated for all $\Phi$.
(Recall that $J$ denotes the Jacobian determinant of $\Phi$.)
To see this, compute
\begin{equation}\label{extend}
\int_\Omega (\xi^i\!\circ\!\Phi)(X) J(X) ((\partial_j \Theta_i{}^j)\!\circ\!\Phi)(X) \,dV(X) = \int_{\Phi(\Omega)}
\xi^i (x)\,(\partial_j \Theta_i{}^j)(x) \,dV(x).
\end{equation}
Now the integral on the right-hand side of (\ref{extend}) can be extended to all of $\mathbb{R}^3$. After partial integration,
using that $\xi$ diverges at most linearly at infinity, there follows
\begin{equation}\label{partial}
\int_{\Omega}\mathscr{F}_i [\Phi](X) dV(X) = - \int_{\mathbb{R}^3} (\partial_j \xi_i)(x) \Theta^{ij}(x) dV(x) = 0
\end{equation}
An example is afforded by the following situation: Let $\rho$ be the function on $\mathbb{R}^3$ given by
\begin{equation}\label{rho}
\rho[\Phi] (x) = [(J\!\circ\! \Phi^{-1})(x)]^{-1} \rho_0\;\chi_{\Phi(\Omega)}
\end{equation}
with $\chi$ be the characteristic function and $\rho_0$ a constant. This is the spatial density corresponding
to the (constant) reference density
$\rho_0$. Consider, then, the function $U$ on $\mathbb{R}^3$ given by
\begin{equation}\label{U}
\Delta U = - 4 \pi G \rho
\end{equation}
with $U (x) = O(|x|^{-1})$ at infinity, i.e. the gravitational potential generated by $\rho$. Now take
\begin{equation}\label{theta}
\Theta_{ij} = \frac{1}{4 \pi G} [(\partial_i U) (\partial_j U) - \frac{1}{2}\delta_{ij}\, \delta^{kl}(\partial_k U) (\partial_l U)].
\end{equation}
Then
\begin{equation}\label{rhodU}
\partial_j \Theta_i{}^j = - \rho \partial_i U
\end{equation}
so that (\ref{forall}) is satisfied. In this case, equlibration can also be checked by first computing
\begin{equation}\label{explicit}
\mathscr{F}_i[\Phi](X) = G \rho_0 \int_\Omega \frac{\Phi_i(X) - \Phi_i (Y)}{|\Phi(X) - \Phi(Y)|^3}\; dV(Y),
\;\;\;\;\;\Phi_i = \delta_{ij} \Phi^j
\end{equation}
and using the explicit form of the Killing vectors. We have of course rediscovered
the well-known fact that the total force and the total torque
on a body due to its own gravitational field is zero.
\section{Circular orbits}
\subsection{Test body in a central gravitational field}
Here we have consider the gravitational field of a point source of mass $M$, in a reference frame
rigidly rotating with angular frequence $\omega$.
This leads to a spatial, harmonic force field given by
\begin{equation}\label{GM}
\bar{K}_i = \rho_0 \;\partial_i \Big[\omega^2 \,\frac{(x^1)^2 + (x^2)^2}{2} + \frac{GM}{r}\Big],
\end{equation}
where $r = \sqrt{(x^1)^2 + (x^2)^2 + (x^3)^2}$. The field $\bar{K}_i$ vanishes on the circle given
by $r = L, x^3 =0$ where
\begin{equation}\label{GML}
\omega^2 L = \frac{GM}{L^2},
\end{equation}
which of course corresponds to circular orbits of point particles. Thus we take $\lambda
\mathscr{F}_i = \lambda K_i \circ \Phi$ with
\begin{equation}\label{centri}
K_i = \partial_i \Big[\frac{(x^1)^2 + (x^2)^2}{2} + \frac{L^3}{r}\Big]
\end{equation}
on $\mathbb{R}^3 \backslash \{0\}$
and solve
\begin{equation}\label{equ1}
E + \lambda F = 0,
\end{equation}
where $E$ and $F$ are respectively defined after Eq.(\ref{defE}) and in Eq.(\ref{load map}). For
small $\lambda$ near the reference state defined by
\begin{equation}\label{reference}
\bar{\Phi} = \mathrm{id}|_\Omega,
\end{equation}
where $\Omega = B_R(l)$. Here $R<L$ and $l$ is for concreteness taken to be $l = (L,0,0)$. (The
body is here and henceforth assumed to have spherical shape in its relaxed configuration.) The
interpretation of a solution $\Phi_\lambda$ for small $\lambda$ will be that of an elastic body
moving with constant angular frequency $\omega$ along a circle (of radius $\sim \Phi^1 (L,0,0)$) in
the central gravitational field of a mass $M$.
 Furthermore there holds
\begin{equation}\label{interpret}
\lambda = \rho_o \omega^2,\;\;\;\;\;\omega^2 = \frac{GM}{L^3}
\end{equation}
Since the function in square brackets in Eq.(\ref{centri}) has Laplacian equal to a 2, it follows from our previous discussion that
$\mathscr{F}$ is equilibrated at $\Phi = \bar{\Phi}$.\\
Our application of the implicit function theorem will be much
simplified if we a priori impose the condition that the
configurations be symmetric with respect to reflection at the planes
$X^2=0$ and $X^3=0$. Thus we now restrict
$W^{2,p}(\Omega,\mathbb{R}^3)$ to the Banach space
$W_{\mathrm{sym}}^{2,p}(\Omega,\mathbb{R}^3)$ of all maps $\Phi \in
W^{2,p}(\Omega,\mathbb{R}^3)$, so that
\begin{eqnarray}\label{sym}
\Phi^1(X^1,X^2,X^3) &=  \Phi^1(X^1,- X^2,X^3) &=\Phi^1(X^1,X^2,- X^3) \notag\\
\Phi^2(X^1,X^2,X^3) &= - \Phi^2(X^1,- X^2,X^3) &= \Phi^2(X^1,X^2,- X^3) \\
\Phi^3(X^1,X^2,X^3) &= \Phi^3(X^1,- X^2,X^3) &= - \Phi^3(X^1,X^2,- X^3) \notag
\end{eqnarray}
These relations are the same as those satisfied by a vector field, which has mirror symmetry w.r.
to both the $X^2=0$ and the $X^3=0$ plane. Since the set (\ref{sym}) contains the identity, there
will again be an open neighbourhood of the identity, which we call $\mathscr{C}_{\mathrm{sym}}$, so
that each element of $\mathscr{C}_{\mathrm{sym}}$ has a $C^1$ - inverse. We will define as
restricted load space the $\mathscr{L}_{\mathrm{sym}}$ given by all $(b,\tau) \in
W^{0,p}(\Omega,\mathbb{R}^3) \times W^{1-1/p,p}(\partial \Omega,\mathbb{R}^3)$, with components
having the same symmetries (\ref{sym}). Supposing that the stored energy function is homogenous and
isotropic, it follows that $\partial_A \sigma_i{}^A$ also satifies (\ref{sym}). Similarly, using
that $\Omega$ - whence $\partial \Omega$ - is invariant under the mirror symmetry, it follows that
$\sigma_i{}^A n_A$ is also invariant. Thus the elasticity operator $E$ maps
$\mathscr{C}_{\mathrm{sym}}$ into $\mathscr{L}_{\mathrm{sym}}$. Next observe that the equilibration
conditions (\ref{equil}) in $\mathscr{C}_{\mathrm{sym}}$ are by symmetry all identically satisfied
except for that where $\xi = \partial_1$. Thus $\mathscr{L}_{\mathrm{sym}}^e$, the set of
equilibrated loads in $\mathscr{L}_{\mathrm{sym}}$, is simply given by all $(b,\tau) \in
\mathscr{L}_{\mathrm{sym}}$, for which
\begin{equation}\label{equilloads}
\int_{B_R(l)}b_1(X) \;dV(X) + \int_{\partial B_R(l)} \tau_1(X) \;dS(X) = 0.
\end{equation}
In particular the set $\mathscr{L}_{\mathrm{sym}}^e$ does not depend on $\Phi$. It is clearly a
Banach subspace of $\mathscr{L}_{\mathrm{sym}}$ of codimension 1. Note next that the only Killing
vectors in $W_{\mathrm{sym}}^{2,p}(\Omega,\mathbb{R}^3)$ are translations in the $X^1$ - direction.
Thus the linearized operator $\delta E$ at $\bar{\Phi}$ has kernel exactly given by the linear span of $\delta \Phi = \partial_1$.\\
Now turn to operator $F$ given by $\Phi \in \mathscr{C}_{\mathrm{sym}} \mapsto (b_i = K_i \circ
\Phi, \tau_i =0) \in \mathscr{L}$. By well known results on ``Nemitskii operators'' (see
\cite{VA}), this map is $C^1$. Furthermore it restricts to a map from $\mathscr{C}_{\mathrm{sym}}$
to $\mathscr{L}_{\mathrm{sym}}$. Now consider the subset $\mathscr{C}_\mathscr{F} \subset
\mathscr{C}_{\mathrm{sym}}$ subject to
\begin{equation}\label{CF}
\mu_{R,l}[\partial_1;\Phi] = \int_{B_R(l)} K_1 \circ \Phi\; d^3 X = 0.
\end{equation}
As shown in Sect.(3.1), the identity map $\bar{\Phi}$ lies in $\mathscr{C}_{\mathscr{F}}$.
Using (\ref{varmean}) and Eq.(\ref{centri}), an easy computation shows that
\begin{equation}\label{delCF}
\delta_\Phi \mu_{R,l}[\partial_1;\bar{\Phi}] \cdot \partial_1 = |B_{R}|\cdot 3
\end{equation}
Thus $\mathscr{C}_\mathscr{F}$ is a $C^1$ - submanifold $\subset \mathscr{C}_{\mathrm{sym}}$ near
$\bar{\Phi}$ of codimension 1 (see e.g. \cite{LA}). Furthermore, by construction, $E + \lambda F$
maps $\mathscr{C}_{\mathscr{F}}$ into $\mathscr{L}_{\mathrm{sym}}^e$ and its linearization at
$\lambda = 0, \Phi = \bar{\Phi}$ is an isomorphism.
The implicit function theorem immediately gives the following \\
{\bf{Theorem 4.1:}} Let the stored energy function entering the elasticity map $E$ be homogenous
and isotropic. Fix a sphere of radius $R$ and a number $L
> R$. Let $\omega$, $\lambda$ be defined by (\ref{interpret}). Then there exists a positive number
$\epsilon$ - which depends on $L,R$ and the stored energy function $\hat{w}$ - so that a solution
$\Phi_\lambda$ of Eq.(\ref{equ1}) exists for $\lambda \in [0,\epsilon)$.

\subsection{Test body with self-gravity}
Now the load $\mathscr{F}_i$ is given by
\begin{equation} \label{plusself}
\lambda \mathscr{F}_i = \lambda \Big[ K_i \circ \Phi +  \frac{\rho_0 L^3}{M} \int_\Omega
\frac{\Phi_i(X) - \Phi_i (Y)}{|\Phi(X) - \Phi(Y)|^3}\; dV(Y)\Big],
\end{equation}
with $K_i$ given by Eq.(\ref{centri}) and $\lambda$ given by (\ref{interpret}).
 Using Sec(3.2) and the fact that (see \cite{BS1}) Eq.(\ref{explicit}) defines a $C^1$ - map
from $\mathscr{C}_{\mathrm{sym}}$ to $\mathscr{L}_{\mathrm{sym}}^e \subset
\mathscr{L}_{\mathrm{sym}}$, everything literally goes through as before and we have the\\
{\bf{Theorem 4.2:}} Under the same assumptions as in Theorem (4.1) the equation $E + \lambda F = 0$
has a unique solution $\Phi_\lambda$ with $\Phi_0 = \bar{\Phi}$, for sufficiently small $\lambda$.
\subsection{Two identical bodies with gravitational interaction and self-gravity}
Here we take for the relaxed state two spheres of radius $R$ at distance $L$ along the $X^1$ axis.
Thus, for the reference configuration we take balls $\Omega = B_R(\frac{1}{2}l)$ and $\Omega' =
B_R(-\frac{1}{2}l)$, where the vector $l$ is again given by $l=(L,0,0)$ with $2R < L$, and
deformations $\Phi:\Omega \rightarrow \mathbb{R}^3$, $\Phi': \Omega' \rightarrow \mathbb{R}^3$ with
\begin{eqnarray}\label{leftright}
\Phi'^1(X^1,X^2,X^3) &=& - \Phi^1(-X^1,X^2,X^3) \notag\\\Phi'^2(X^1,X^2,X^3) &=& \Phi^2(-X^1,X^2,X^3)\\
\Phi'^3(X^1,X^2,X^3) &=& \Phi^3(-X^1,X^2,X^3)\notag
\end{eqnarray}
Thus the relaxed bodies and their deformations have complete mirror
symmetry with respect to the plane $X^1 = 0$ (see Fig.1).

\begin{figure}
\psfrag{X1}[cc][cc]{$X^1$} \psfrag{X3}[cc][cc]{$X^3$}
\psfrag{L}[bc][tc]{$L$}
 \centering
\includegraphics[width=0.5\textwidth]{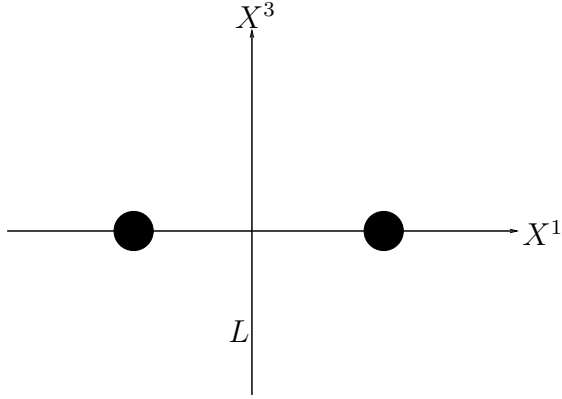}
\caption{Two spheres in steady rotation}
\end{figure}

The force field is given by
\begin{align}\label{combined}
\lambda \mathscr{F}_i[\Phi](X) &= \rho_0 \omega^2
\Big[\delta_i{}^\alpha \Phi_\alpha(X) -
\frac{L^3}{2 |B_R|}\int_{B_R(-\frac{l}{2}l)} \frac{\Phi_i(X) - \Phi'_i (Y')}{|\Phi(X) - \Phi'(Y')|^3}\; dV(Y')\, - \notag\\
& - \frac{L^3}{2 |B_R|}\int_{B_R(\frac{1}{2}l)}
\frac{\Phi_i(X) - \Phi_i (Y)}{|\Phi(X) - \Phi(Y)|^3}\; dV(Y)\Big]
\end{align}
where $\alpha=1,2$. The interpretation of a solution $\Phi_\lambda$ is that of a pair of elastic
bodies of mass $m= |B_R|\, \rho_0$ (reduced mass $M_r$), rotating about their centre of mass with
angular frequency $\omega$  where
\begin{equation}\label{reduced}
\lambda = \rho_0 \omega^2,\;\;\;\;\omega^2 L = \frac{G
M_r}{L^2},\;\;\;\;M_r = \frac{m}{2}\,.
\end{equation}
The self-interaction term in (\ref{combined}) is again equilibrated for all $\Phi$'s. The first two
terms in (\ref{combined}) are again of the form $K_i \circ \Phi$ with $K_i$ a harmonic force field,
where however, for the second term, the potential for this force field is in turn a functional $K$
of $\Phi'$, namely there holds
\begin{equation}\label{K2}
\Delta K [\Phi'](x) = - 4 \pi \rho,
\end{equation}
where (see (\ref{rho}))
\begin{equation}
\rho[\Phi'] (x) = [(J'\!\circ\! \Phi'^{-1})(x)]^{-1} \rho_0\;\chi_{\Phi'(\Omega)}
\end{equation}
with $J'$ the Jacobian of the map $\Phi'$.
We again find that $\mathscr{F}$ is equilibrated at the identity. Furthermore
\begin{equation}
\delta_\Phi \mu_{R,\frac{1}{2}l}[\partial_1;\bar{\Phi}] \cdot \partial_1 = |B_R| \cdot 3
\end{equation}
for the sum of forces in (\ref{combined}), of course the last term does not contribute to $\mu$.
Thus exactly as in the previous section $\mathscr{C}_\mathscr{F}$ is a $C^1$ - submanifold $\subset
\mathscr{C}_{\mathrm{sym}}$ near $\bar{\Phi}$ of codimension 1 and $E + \lambda F$ maps
$\mathscr{C}_{\mathscr{F}}$ into $\mathscr{L}_{\mathrm{sym}}^e$ with its linearization at $\lambda
= 0, \Phi =
\bar{\Phi}$ an isomorphism. We have obtained the \\
{\bf{Theorem 4.3:}} Consider Eq.(\ref{equ}) with homogenous and isotropic stored energy function
and with $F$ of the form $F:\Phi \in \mathscr{C}' \mapsto (b_i = \mathscr{F}_i[\Phi],\tau_i = 0)$
where $\mathscr{F}_i$ is given by Eq.(\ref{combined}). For sufficiently small $\lambda$ there
exists a solution $\Phi_\lambda$ with $\Phi_0 = \bar{\Phi}$.
\section{Static problems}
In Newtonian theory it seems obvious that there exist no static solutions with two gravitating
bodies: the bodies should fall onto each other. This however is in general only true if the
geometry is such that the bodies are separated by a plane. In this section we show that for more
involved geometries there do exist static 2--body solutions.

First we try to put a small body near a critical point of the field of some external body.  The
gravitational force vanishes at critical points of the gravitational potential due to this body. In
section (5.1) we will simply assume that a critical point, subject to some further restrictions, be
given, and then show that there exists an equilibrium configuration of a small elastic body near
that critical point. Such critical points are however absent in those parts of the vacuum region
which are separated by a plane from the support of the large body. One thus has to put the small
body in the "hollow space" exterior to a sufficiently non-convex body (Fig.2a). We give a
particular example for such a body in section (5.2) corresponding to Fig.2b, where we also consider
the full 2 body problem.

\begin{figure}
\centering
\subfigure[]{\label{cavity}\includegraphics[width=0.4\textwidth]{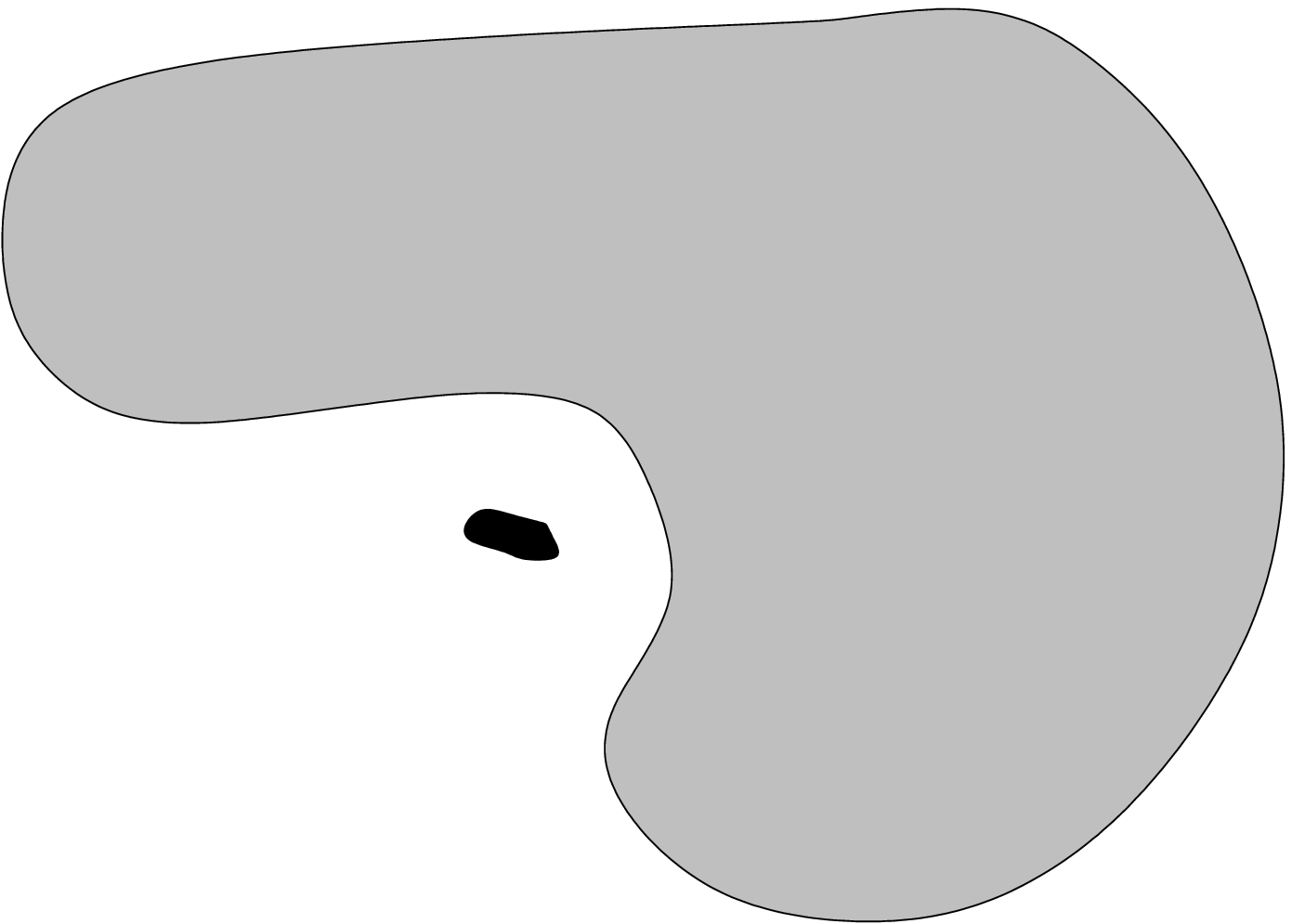}}
\hspace*{5em}
\subfigure[]{\includegraphics[width=0.25\textwidth]{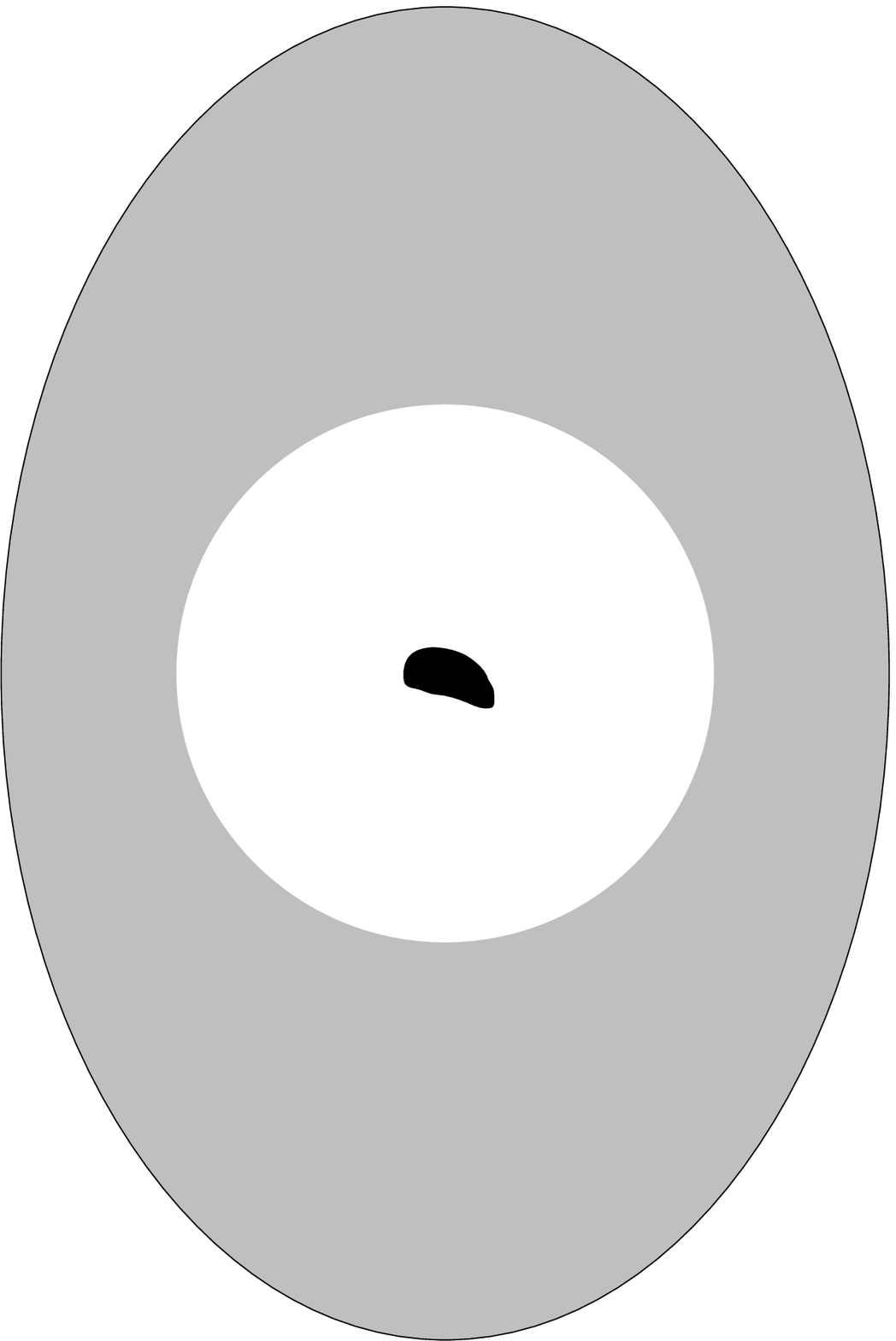}}
\caption{static configurations}\label{cavityellipsoid}
\end{figure}

\subsection{Test body}
We first treat the case of a body in a given potential field $K_i = \partial_i K$. We make the following assumptions:\\
\begin{enumerate}
\item $K$ has a nondegenerate critical point at the origin. \item The domain $\Omega$ has the
origin as centre of mass. \item The Hessian of $K$ and the tensor of inertia of $\Omega$ are
simultaneously diagonalizable (i.e. commute). \item The eigenvalues of the inertia tensor are all
different, and the eigenvalues of the Hessian are all different.
\end{enumerate}
We now show that, by simultaneously translating, rotating and scaling $\Omega$, one can satisfy the
equilibration conditions (at the identity). Let $f_{ij}=f_{(ij)}$ be the Hessian of $K$ at the
origin, We first observe that the force $\mathring{K}_i = f_{ij} x^j$ is equilibrated w.r. to
$\Omega$. Namely
\begin{equation} \label{inforce}
\mathring{\mu}_\Omega [d^i \partial_i;\bar{\Phi}] = \int_\Omega d^i
f_{ij} x^j dV(x) = 0,
\end{equation}
by Ass.2 and
\begin{equation} \label{inmom}
\mathring{\mu}_\Omega [\omega^i{}_j x^j \partial_i;\bar{\Phi}] =
\omega^{ij} f_{ik} \Theta_j{}^k,
\end{equation}
where
\begin{equation} \label{inert}
\Theta^{jk} = \int_\Omega x^j x^k d V(x).
\end{equation}
Expression (\ref{inmom}) is zero on account of $\omega_{ij}=\omega_{[ij]}$ together with Ass.3.
(The quantity $\Theta^{ij}$ differs from the standard definition of the inertia tensor by a
multiple of $\delta^{ij}$, so properties 3 and 4 carry over to $\Theta^{ij}$.) The full force $K_i$
can be written as
\begin{equation}\label{full}
K_i = f_{ij} x^j + f_{ijk}(x)x^j x^k.
\end{equation}
Denote by $\bar{\Phi}_{S,D}$ the map $x^i \mapsto \bar{x}^i =
S^i{}_j x^j + D^i$ with $S$ a rotation matrix and consider the
expression
\begin{equation}\label{K}
k(d,\epsilon;S,D) = \frac{1}{\epsilon^4}\,
\mu_{\Omega_\epsilon}[d^i\partial_i;\bar{\Phi}_{S,\,\epsilon D}],
\end{equation}
where $\Omega_\epsilon = \{x \in \mathbb{R}^3|\frac{1}{\epsilon} x
\in \Omega\}$. After the obvious change of integration variable it
follows that
\begin{eqnarray}\label{K1}
\lefteqn{ k(d,\epsilon;S,D) = \int_\Omega d^i [f_{ij}(S^j{}_{j'} y^{j'} +
D^j) + {}} \nonumber\\
&&{}+ \epsilon\, f_{ijk}(\epsilon S y + \epsilon D)(S^j{}_{j'}
y^{j'} + D^j) (S^k{}_{k'} y^{k'} + D^k)] dV(y).
\end{eqnarray}
Similarly, the quantity
\begin{equation}\label{M}
m(\omega,\epsilon;S,D)= \frac{1}{\epsilon^5}\, \mu_{\Omega_\epsilon}[\omega^i{}_j x^j
\partial_i;\bar{\Phi}_{S,\,\epsilon D}]
\end{equation}
is equal to
\begin{eqnarray}\label{M1}
\lefteqn{ m (\omega,\epsilon;S,D) = \int_\Omega \omega^i{}_j(S^j{}_{j'} y^{j'} + D^j)
[f_{ik}(S^k{}_{k'}
y^{k'} + D^k) + {}} \nonumber\\
&&{}+ \epsilon\, f_{ikl}(\epsilon S y + \epsilon D)(S^k{}_{k'}
y^{k'} + D^k) (S^l{}_{l'} y^{l'} + D^l)] dV (y).
\end{eqnarray}
Clearly we have that $k(d,0;\mathrm{id},0)=0,\,m(\omega,0;\mathrm{id},0)= 0$. We can view the pair
$(k(\,\cdot\,,\epsilon,S,D),m (\,\cdot\,,\epsilon,S,D))$ as defining an $\epsilon$ - dependent map
from the Euclidean group $\mathbb{E}(3)$ into the dual of the Lie algebra of $\mathbb{E}(3)$, and
the derivative of this map as a bilinear form $V$ on the Lie algebra of $\mathbb{E}(3)$. We find
that
\begin{eqnarray}\label{var1}
\delta_D k(d,0\,;\mathrm{id},0) \cdot e &=& |\Omega|\; d^i f_{ij}e^j \\
\label{var2}
\delta_S k(d,0\,;\mathrm{id},0) \cdot \mu &=& 0 \\
\label{var3}
\delta_D m(\omega,0\,;\mathrm{id},0) \cdot e &=& 0 \\
\label{var4} \delta_S m(\omega,0\,;\mathrm{id},0) \cdot \mu &=& |\Omega|\; (f_{ij}\Theta^{kl} +
f_{mi} \Theta^{ml} \delta^k{}_j) \omega^i{}_k \mu^j{}_l\;,
\end{eqnarray}
where $\mu_{ij} = \mu_{[ij]}$ and we have used Ass.2 in (\ref{var2}) and (\ref{var3}). The
expression in brackets on the r.h. side of (\ref{var4}), viewed as a bilinear form on antisymmetric
tensors, is symmetric by Ass.3. Writing $\omega^i{}_j = \epsilon^i{}_{jk} p^k,\;\mu^i{}_j =
\epsilon^i{}_{jk} q^k$, this form is equivalent to $|\Omega|\,\tau_{ij}\,p^i q^j$ where
$\tau^i{}_j$, by the standard $\epsilon$ - identities and using Ass.3, is in matrix notation given
by
\begin{equation}\label{tau}
\tau = 3\, \Theta \,f - (\mathrm{tr}\,f) \Theta - (\mathrm{tr}
\,\Theta) f + [(\mathrm{tr}\, \Theta)(\mathrm{tr}\, f) - 2
\,\mathrm{tr}\,(\Theta \,f)]\, \mathrm{id}
\end{equation}
Going to a frame where both $\Theta$ and $f$ are diagonal, we see that $\tau$ is also diagonal,
with e.g. $\tau_{11} = (f_{33} - f_{22})(\Theta_{22} - \Theta_{33})$. Thus $\tau$ is
non-degenerate, using Ass.4. Using Ass.1, the full bilinear form $V$ is also non-degenerate.
Therefore by the implicit function theorem we conclude that there exists, for $\epsilon$
sufficiently close to 0, a unique family $(S(\epsilon),D(\epsilon))$ with $(S(0),D(0)) =
(\mathrm{id},0)$ so that $\mu_{\Omega_\epsilon}[d^i\partial_i;\bar{\Phi}_{S (\epsilon),\,\epsilon
D(\epsilon)}]=0$ for all $d^i$ and $\mu_{\Omega_\epsilon}[\omega^i{}_j x^j
\partial_i;\bar{\Phi}_{S(\epsilon),\,\epsilon D(\epsilon)}]=0$ for all $\omega^i{}_j$
with $\omega_{ij} = \omega_{[ij]}$.\\
We next define $\bar{\Omega} = \Omega_\epsilon$ for fixed small positive $\epsilon$ and, by slight
abuse of notation, $\bar{\Phi} = \mathrm{id}_{\bar{\Omega}}$. Again our task is to solve
(\ref{equ}), namely
\begin{equation}\label{equ'}
E[\Phi] + \lambda F[\Phi] = 0\;,
\end{equation}
where the load map $F$ is of the form
\begin{equation}
F:\Phi \in \mathscr{C} \mapsto (b_i = \mathscr{F}_i[\Phi],\tau_i =
0)
\end{equation}
and $\mathscr{F}_i [\Phi] = K_i \circ \Phi$. Both $E$ and $F$ are viewed as maps from $\mathscr{C}$
to the load space $\mathcal{L}$ (see Sect.2). We have the following\\
{\bf{Theorem 5.1}}: For sufficiently small $\lambda$ equation (\ref{equ'}) has a unique solution
$\Phi_\lambda \in \mathscr{C}$ near
$\bar{\Phi}$.\\
{\bf{Proof}}: In a similar way as in Sect.4.1 we define $\mathscr{C}_{\mathscr{F}}$ as a
neighborhood of $\bar{\Phi}$ in $\mathscr{C}$ satisfying the 6 conditions
\begin{equation}\label{mean'}
\mu_{\bar{\Omega}}[\xi;\Phi] = \int_{\bar{\Omega}} (\xi^i\!\circ\!
\Phi)(X)\; \mathscr{F}_i [\Phi](X) \;dV(X) = 0
\end{equation}
for all Euclidean Killing vectors $\xi$. Consider the expression
\begin{equation}\label{varmean'}
\delta_\Phi \mu_{\bar{\Omega}}[\xi;\bar{\Phi}] \cdot \delta \Phi =
\int_{\bar{\Omega}} (\mathcal{L}_\xi F)_i \,\delta \Phi^i\, dV(X)
\end{equation}
If $\delta \Phi^i$ is any non-zero Killing vector $\eta$ and $\epsilon$ entering the definition of
$\bar{\Omega}$ is sufficiently small, it follows from the above discussion and continuity that the
expression (\ref{varmean'}) can not be zero for all $\xi$. Thus $\mathscr{C}_{\mathscr{F}}$ is a
$C^1$ - submanifold of
$\mathscr{C}$ of codimension 6.\\
There is an operator $P:\mathscr{L} \rightarrow
\mathscr{L}_{\bar{\Phi}}$ projecting onto loads equilibrated at the
identity. Such an operator is constructed by choosing a (6 -
dimensional) complement of $\mathscr{L}_{\bar{\Phi}}$ in
$\mathscr{L}$ and requiring this complement to be the kernel of $P$.
We now consider the modified equations
\begin{equation}\label{mod}
P \circ (E + \lambda F) = 0.
\end{equation}
for $\Phi \in \mathscr{C}_{\mathscr{F}}$. The linearization $\delta E : W^{2,p}(\bar{\Omega})
\rightarrow \mathscr{L}_{\bar{\Phi}}$ at $\lambda = 0$ of the map in Eq.(\ref{mod}) is clearly an
isomorphism, so there exists, for small $\lambda$, a solution $\tilde{\Phi}$ of Eq.(\ref{mod}) near
$\bar{\Phi}$ $\in \mathscr{C}_{\mathscr{F}}$. By property 2 of the elasticity operator in Sect.2 we
have that, for all $\Phi \in \mathscr{C}$, $E [\Phi] \in \mathscr{L}_{\Phi}$, the set of all loads
satisfying the equilibration conditions at $\Phi = \tilde{\Phi}$, so this holds trivially for
$\tilde{\Phi}$. We also have that $F[\tilde{\Phi}] \in \mathscr{L}_{\tilde{\Phi}}$, since
$\tilde{\Phi} \in \mathscr{C}_{\mathscr{F}}$. Thus $(E + \lambda F)[\tilde{\Phi}] \in
\mathscr{L}_{\tilde{\Phi}}$. By Eq.(\ref{mod}) the load $(E + \lambda F)[\tilde{\Phi}]$ lies in
some complement of $\mathscr{L}_{\bar{\Phi}}$. This complement, by continuity, is also a complement
of $\mathscr{L}_{\tilde{\Phi}}$. Thus $(E + \lambda F)[\tilde{\Phi}] = 0$, and the proof is
complete.
\subsection{A static 2-body problem}
We first need a domain  whose Newtonian potential (assuming constant density) has a critical point
in the vacuum region with all three eigenvalues of the Hessian non-zero and different.
%Then,
%clearly, a small domain  can be constructed near this critical point satisfying Assumptions 1 to 4,
%and from this, by the result of the previous subsection, a domain $\Omega$ so that
%$\mu_{\Omega}[\xi;\bar{\Phi}] = 0$ for all Euclidean Killing vectors
%$\xi$.
\\
Consider first the Newtonian potential $U_0$ of a solid ellipsoid centered at the origin with axes
$a_1 > a_2 > a_3$ and constant density $\rho'$. From chapter 3 of \cite{CH} (Eq.'s (40) and (18))
we read off that that
\begin{equation}\label{solid}
(\partial_i \partial_j U_0) (0) = - 2 \pi G \rho'\, \mathrm{diag}(A_1,A_2,A_3),
\end{equation}
where $0 < A_1 < A_2 < A_3$. In the limit when $a_3$ approaches
$a_2$ we find (using either Eq.(18) of \cite{CH} or the explicit
expressions (38,39) of \cite{CH}) that $A_1 < A < A_2 = A_3$, where
$A$ is the value of $A_i$ when $a_1 = a_2 = a_3$ (which is equal to
$\frac{2}{3}$). By continuity we have $A_1 < A < A_2 < A_3$ for
$a_3$ less than but close to $a_2$. We now subtract from $U_0$ the
potential of a solid sphere of radius $a < a_3$, centered at the
origin, thus obtaining a new potential $U$ corresponding to a hollow
triaxial ellipsoid $\Omega'$. The potential $U$ has a critical point
at
the origin with Hessian having one positive and two negative eigenvalues, all different.\\
We now choose a domain $\Omega$ with mass center at this critical point and such that the inertia
tensor satisfier Ass.'s 3 and 4 in Sect. 5.1: as explained in Sect. 5.1 we can, by translating,
rotating and scaling $\Omega$, find a domain $\bar{\Omega}$ so that
$\mu_{\bar{\Omega}}[\xi;\mathrm{id}_{\bar{\Omega}}] = 0$ for all Euclidean Killing vectors $\xi$.
By slight abuse of notation we henceforth call $\bar{\Omega}$ again $\Omega$. Using an argument
identical to that in section (3.2), one sees that when $\mu_{\Omega}[\xi;\mathrm{id}_\Omega] = 0$
there also follows that $\mu_{\Omega'}[\xi;\mathrm{id}_{\Omega'}] = 0$ for all Euclidean Killing
vectors ("actio = reactio") - and this can also be checked explicitly. We now assign to the two
domains $\Omega'$ and $\Omega$ the constant densities $\rho'$ and $\rho$ and elasticity operators
$E'$ mapping $\Phi' \in \mathscr{C}'$ into the load space $\mathscr{L}'$ and $E$, mapping $\Phi
\in\mathscr{C}$ into the load space $\mathscr{L}$, respectively, with both stored energy functions
satisfying Eq.'s (\ref{stressfree},\ref{pointwise}). Neither isotropy nor homogeneity of the stored
energy functions is needed in this section. Associated with $\Omega'$ there is the load map $F'$
with $\mathscr{F}'$ given by
\begin{align}\label{loadprime}
\lambda \mathscr{F}_i' [\Phi,\Phi'](X') &= - G \Big[\rho \int_{\Omega} \frac{\Phi_i'(X') - \Phi_i
(Y)}{|\Phi'(X') - \Phi(Y)|^3}\; dV(Y)\, +\notag\\ &+ \rho' \int_{\Omega'} \frac{\Phi'_i(X') -
\Phi'_i (Y')}{|\Phi'(X') - \Phi'(Y')|^3}\; dV(Y')\Big]
\end{align}
and with $\Omega$ there is associated $\mathscr{F}$ given by
\begin{align}\label{loadunprime}
\lambda \mathscr{F}_i [\Phi,\Phi'](X) &= - G \Big[\rho' \int_{\Omega'} \frac{\Phi_i(X) - \Phi'_i
(Y')}{|\Phi(X) - \Phi'(Y')|^3}\; dV(Y')\, +\notag\\ &+ \rho \int_{\Omega} \frac{\Phi_i(X) - \Phi_i
(Y)}{|\Phi(X) - \Phi(Y)|^3}\; dV(Y)\Big].
\end{align}
We now define $\mathscr{D}_{\mathscr{F}} \subset \mathscr{C} \times \mathscr{C}' \subset
W^{2,p}(\Omega,\mathbb{R}^3) \times W^{2,p}(\Omega',\mathbb{R}^3),\;p > 3$ as the set of
$(\Phi,\Phi') \in \mathscr{C} \times \mathscr{C}'$ for which
\begin{equation}\label{equilprimeunprime}
\int_{\Omega \times \Omega'} \xi^i (X)\, \frac{\Phi_i(X) - \Phi'_i (Y')}{|\Phi(X) - \Phi'(Y')|^3}\;
dV(X)\, dV(Y') = 0
\end{equation}
and $\Phi'$ satifies
\begin{equation}\label{fix}
\Phi'_i (X_0) = 0,\;\;\partial'_{[i}\Phi'_{j]}(X_0) = 0,
\end{equation}
where $X_0$ is some given point in $\Omega'$. Again, by section (3.2) or by calculation using the
explicit form of $\xi^i (X)$, one shows that Eq.(\ref{equilprimeunprime}) is equivalent to
\begin{equation}\label{equilunprimeprime}
\int_{\Omega \times \Omega'} \xi^i (X')\, \frac{\Phi'_i(X') - \Phi_i (Y)}{|\Phi'(X') -
\Phi(Y)|^3}\; dV(X)\, dV(Y') = 0
\end{equation}
Combining the fact that there are no non-zero Killing vectors vanishing at a point together with
their curl and the discussion of the previous subsection - making $\epsilon$ entering the
definition of $\Omega$ smaller if necessary, it follows that $\mathscr{D}_{\mathscr{F}}$ is a $C^1$
- submanifold of $\mathscr{C} \times \mathscr{C}'$ near $(\bar{\Phi},\bar{\Phi}')$ of codimension
12. We now set $\lambda = G$ and solve, using the implicit function theorem, the equations
\begin{equation}
P  ( E[\Phi] + \lambda F[\Phi,\Phi']) =0,\hspace{0.5cm}P'(E'[\Phi'] + \lambda F'[\Phi,\Phi']) = 0
\end{equation}
on $\mathscr{D}_{\mathscr{F}}$, for small $G$ and $(\Phi,\Phi')$ near $(\bar{\Phi},\bar{\Phi}')$.
Here $P$, $P'$ are projection operators on $\mathscr{L}$ resp. $\mathscr{L}'$, defined analogously
as in Sect.(5.1). The argument showing that we have in fact also solved the "unprojected" equations
proceeds as in Sect.(5.1), if it is recalled that the self force terms in
(\ref{loadunprime},\ref{loadprime}) are identically equilibrated. We thus have the\\
{\bf{Theorem 5.2:}} Consider the coupled set of equations
\begin{equation}\label{coupled}
E[\Phi] + \lambda F[\Phi,\Phi'] =0,\hspace{0.5cm}E'[\Phi'] + \lambda F'[\Phi,\Phi'] = 0,
\end{equation}
where $(\Phi,\Phi') \in \mathscr{C} \times \mathscr{C}'$. Then, for $\lambda$ sufficiently small,
there is a unique solution $(\Phi_\lambda,\Phi'_\lambda)$ with $(\Phi_0,\Phi'_0) =
(\mathrm{id}_\Omega,\mathrm{id}_{\Omega'})$.


\begin{thebibliography}{99}
\bibitem[1] {BSR} {Beig, R., Schmidt, B.G.,}  Relativistic Elastostatics. I. Bodies in rigid rotation,
{\it Class. Quantum Grav} {\bf 22},
2249-2268 (2005)
\bibitem[2] {BS1} {Beig, R., Schmidt, B.G.,}  Static, self-gravitating elastic bodies, {\it Proc.R.Soc.London}
{\bf A459} 109 -115 (2003)
\bibitem[3] {CH} {Chandrasekhar, S.,} Ellipsoidal Figures of
Equilibrium, Yale University Press (1969)
\bibitem[4] {CI} {Ciarlet, P.G.,}  Mathematical Elasticity, Volume 1: Three-Dimensional Elasticity,  North-Holland (1988)
\bibitem[5] {GT} {Gilbarg, D., Trudinger, N.S.,} Elliptic partial differential equations of second order,
Grundlehren der Mathematischen Wissenschaften, vol.224, Springer (1983)
\bibitem[6] {LA} {Lang, S.,} Differential Manifolds, Springer (1985)
\bibitem[7] {LI} {Lichtenstein, L.,} Gleichgewichtsfiguren rotierender Fl\"ussigkeiten, Springer  (1933)
\bibitem[8] {MH} {Marsden, J.E., Hughes, T.J.R.,}  Mathematical foundations of elasticity, Dover (1994)
\bibitem[9] {VA} {Valent, T.,}  Boundary Value Problems of Finite Elasticity, Springer (1987)
\end{thebibliography}
\end{document}